\documentstyle[aps,prl,twocolumn,floats,psfig,epsfig]{revtex}

\def\ub{{\overline{u}}}
\def\vb{{\overline{v}}}

\begin{document}
\draft
\twocolumn[\hsize\textwidth\columnwidth\hsize\csname@twocolumnfalse\endcsname
\title{
Critical behavior of the two-dimensional $N$-component Landau-Ginzburg Hamiltonian with cubic anisotropy.
}
\author{Pasquale Calabrese$\,^1$ and Alessio Celi$\,^2$}
\address{$^1$  Scuola Normale Superiore and INFN,
Piazza dei Cavalieri 7, I-56126 Pisa, Italy. 
}
\address{$^2$ 
Dipartimento di Fisica dell'Universit\`a di Milano
and INFN, 
Via  Celoria 16, I-20133 Milano, Italy.
\\
{\bf e-mail: \rm
{\tt calabres@df.unipi.it},
{\tt celi@mi.infn.it}
}}

\date{\today}

\maketitle

\begin{abstract}
We study the two-dimensional $N$-component Landau-Ginzburg Hamiltonian with 
cubic 
anisotropy. We compute and analyze the fixed-dimension perturbative expansion
of the renormalization-group functions to four loops.
The relations of these models with $N$-color Ashkin-Teller models, discrete 
cubic models, planar model with fourth order anisotropy,
and structural phase transition in adsorbed monolayers are discussed.  
Our results for $N=2$~($XY$ model with cubic anisotropy) are compatible with
the existence of a line of fixed points joining the Ising and the $O(2)$ fixed
points. Along this line the exponent $\eta$ has the constant value $1/4$, 
while the exponent $\nu$ runs in a continuous and monotonic way from 1 to 
$\infty$~(from Ising to $O(2)$).
For $N\geq 3$ we find a cubic fixed point in the region
$u ,v \geq 0$, which is marginally stable or unstable according to the
sign of the perturbation.
For the physical relevant case of $N=3$ we find the exponents $\eta=0.17(8)$ 
and $\nu=1.3(3)$ at the cubic transition.   
\end{abstract}

]


\section{Introduction}
\label{introduction}

In the framework of Renormalization Group~(RG) approach to critical phenomena,
the critical behavior at many continuous phase transitions can be investigated 
by considering an effective Landau-Ginzburg Hamiltonian, having an
$N$-component order parameter $\phi_i$ as fundamental field, and containing 
up to fourth-order powers of the field components. The fourth-degree 
polynomial form of the potential depends essentially on the symmetry of 
the system. In fact 
according to the universality hypothesis, the critical properties of
these systems can be 
described in terms of quantities that do not depend on the microscopic details
of the  system, but only on global properties such as the dimensionality
and the symmetry of the order parameter, and the range of the interactions.

The critical properties of many magnetic materials are 
computed using the $O(N)$-invariant Landau-Ginzburg Hamiltonian.
Uniaxial ferromagnets should be described by the Ising universality
class ($N=1$), while magnets with easy-plane anisotropy should belong to
the $XY$ universality class. Ferromagnets are often
described in terms of the $O(3)\,$ Hamiltonian. However, this is correct 
if the non-rotationally invariant interactions that have only the 
reduced symmetry of the lattice are irrelevant in the renormalization-group
sense. 
In two dimensions the effect of anisotropy is very important:
systems possessing continuous symmetry do not exhibit conventional long-range
order at finite temperature, while models with discrete symmetry do undergo
phase transitions into conventionally ordered phase.

For studying the effect of cubic anisotropies one usually consider the 
$\phi^4$ theory \cite{Aharony-76,rev-01}:
\begin{eqnarray}
{\cal H} = \int d^d x &&
\left\{ {1\over 2} \sum_{i=1}^{N}
      \left[ (\partial_\mu \phi_i)^2 +  r \phi_i^2 \right]  \right. \nonumber\\
&& +\left. 
 {1\over 4!} \sum_{i,j=1}^N \left( u_0 + v_0 \delta_{ij} \right)
\phi^2_i \phi^2_j \right\}, 
\label{Hphi4}
\end{eqnarray}
in which the added cubic term breaks explicitly the $O(N$) invariance 
leaving a residual discrete cubic symmetry given by the reflections
and permutations of the field components \cite{footnote0}.  
This term favors the spin orientations towards the faces or the corners of
an $N$-dimensional hypercube for $v_0<0$ or $v_0>0$ respectively.

The Hamiltonian (\ref{Hphi4}) has  received much attention also because
it describes the two-dimensional order-disorder transition in some adsorbed 
monolayers. In fact, in the original classification \cite{dswg-78} on 
square and rectangular lattices, all these transitions belong to the 
universality class of the Ising model, three- or four-state Potts model, and
$XY$ model with cubic anisotropy. In the successive extension \cite{ds-79} to 
non-Bravais lattices, as the honeycomb and the Kagom\'e ones, also the 
Heisenberg~($N=3$) model with cubic anisotropy is interested by this
classification~(e.g. in the adsorption of diatomic-molecules on graphite).
The full classification of continuous phase transition of magnetic symmetry in
two dimensions \cite{class} reveals the interest of the model  (\ref{Hphi4})
also for other kind of transitions.

An important application of the $XY$ model with cubic anisotropy
is in the oxygen ordering in $\mbox{YBaCu}_3 \mbox{O}_{6+x}$ \cite{bew-89,anrl-90},
since it is one of the most studied high-temperature superconductors.
Some other applications of Landau-Ginzburg Hamiltonian (\ref{Hphi4}) are:
the buckling instabilities of a confined colloid crystal layer
\cite{cn-93}, some discrete models with competing nearest- and 
next-nearest-neighbor interactions \cite{ssm-97}, and, 
very recently \cite{bbm-00}, it is argued that 
the phase diagram of a lattice-gas model
for studying the micellar binary solution of water and ampliphile is 
partially described by  the $XY$ model with cubic anisotropy.

We also mention that, in
the limit $N\to0$, the cubic model (\ref{Hphi4}) describes 
the Ising model with site-diluted disorder
\cite{Harris-Lubensky_74,Lubensky_75,Khmelnitskii_75}, but we do not 
consider it here, since in the framework of fixed dimension $d=2$ it was 
already analyzed in Ref. \cite{mayer-89}.

In this paper we present a field-theoretic study based 
on an expansion performed directly in 
two dimensions, as proposed for the $O(N)$ models by Parisi\cite{Parisi-80}. 

The paper is organized as follows. In Sec. II we give an overview of all 
known results that we believe necessary for a full understanding of the 
two-dimensional cubic model (\ref{Hphi4}). 
In Sec. III we derive the perturbative 
series for the renormalization-group functions at four loops and discuss the 
singularities of the Borel transform. The results of the analysis are 
presented in Sec. IV. 
The reader that is not interested to the detail of the calculations can skip
sections III and IV and read directly Sec. V where we summarize all our results
and point out some questions which we think deserve further study.

\section{Overview of Known results}

\subsection{Three dimensional results}

We shortly review the most interesting features
appearing in the study of the Hamiltonian (\ref{Hphi4}) 
in the framework of $\epsilon$-expansion\cite{Aharony-73,Aharony-76} 
and at fixed dimension $d=3$\cite{cpv-99}, since they are necessary for a
good understanding of the two-dimensional case. 

The model (\ref{Hphi4}) has four fixed points:
the trivial Gaussian one, the Ising one in which the $N$ components of the 
field decouple,
the O($N$)-symmetric and the  cubic fixed points.
The Gaussian fixed point is always unstable, and so is 
the Ising fixed point~\cite{Sak-74}.
Indeed, in the latter case, it is natural to interpret Eq. (\ref{Hphi4}) 
as the Hamiltonian of $N$ Ising-like systems coupled by the 
$O(N)$-symmetric term. But this interaction is the sum of the products 
of the energy operators of the different Ising systems.
Therefore, at the Ising fixed point, the crossover exponent 
associated to the O($N$)-symmetric quartic term should be given by
the specific-heat critical exponent $\alpha_I$ of the Ising model,
independently of $N$. 
Since $\alpha_I$ is positive for all $d>2$ 
the Ising fixed point is unstable. Obviously in two dimensions this argument
fails since $\alpha_I =0$.

\begin{figure}[tb]
\vspace{-1.5cm}
\centerline{\psfig{width=8truecm,angle=-90,file=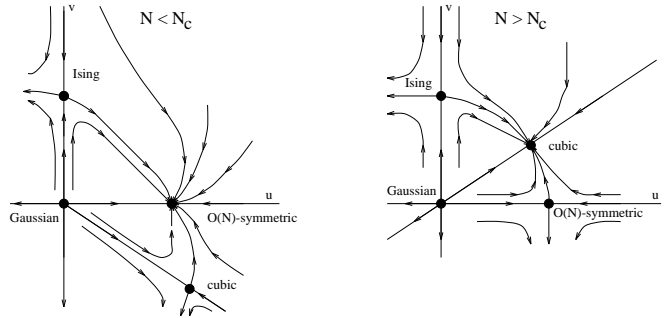}}
\vspace{-1.5cm}
\caption{
Renormalization-group flow in the coupling plane $(u,v)$ for
$N<N_c$ and $N>N_c$.}
\label{rgflow}
\end{figure}

While the  Gaussian and the Ising fixed points are unstable
for any number of components $N$,
the stability properties
of the O($N$)-symmetric and of the cubic fixed points depend on $N$.
For sufficiently small values of $N$, $N<N_c$, the 
O($N$)-symmetric fixed point is stable and the cubic one is unstable.
For $N>N_c$, the opposite is true:
the renormalization-group flow is driven towards the cubic fixed point,
which now describes the generic critical behavior of the system.
The O($N$)-symmetric point corresponds to a tricritical transition.
Figure~\ref{rgflow} sketches the 
flow diagram in the two cases $N<N_c$ and $N>N_c$.
At $N=N_c$, the two fixed points should coincide, and
logarithmic corrections to the O($N$)-symmetric critical exponents are 
expected.
Outside the attraction domain of the fixed points, the flow goes away
towards more negative values of $u$ and/or $v$ and finally
reaches the region where the quartic interaction no longer 
satisfies the stability
condition. These trajectories should be related to
first-order phase transitions \cite{fifopt}.
Some recent and very accurate calculations 
\cite{Shpot-89,M-S-S-89,kleinertetal,S-A-S-97,varn-98,P-S-00,fhy-00,cpv-99} 
suggest that in three-dimensions  $N_c<3$. 

If $N>N_c$, the cubic anisotropy is relevant and therefore the critical
behavior of the system is not described by the Heisenberg isotropic 
Hamiltonian. If the cubic interaction favors the alignment of 
the spins along the diagonals of the cube,
i.e. for a positive coupling $v_0$,
the critical behavior is controlled by the cubic
fixed point and the cubic symmetry is retained even at the 
critical point. On the other hand, if the system tends to 
magnetize along the cubic axes --- this corresponds to a negative
coupling $v_0$ --- then the system undergoes a first-order phase 
transition \cite{Aharony-76,Aharony_77,Wallace-73,S-D-99}.

In the limit $N\to\infty$, keeping $Nu$ and $v$ fixed, one can derive 
exact expressions for the exponents at the cubic fixed point for all dimensions.
Indeed, in this limit the model can be reinterpreted as 
a constrained Ising model~\cite{Emery-75}, leading to a 
Fisher renormalization of the Ising critical exponents~\cite{Fisher-68}.
One has~\cite{Aharony-73-2,Emery-75,Aharony-76}: 
\begin{equation}
\eta = \eta_I+O\left( {1\over N}\right),
\qquad\qquad   \nu = {\nu_I\over 1-\alpha_I}+O\left( {1\over N}\right),
\label{largen}
\end{equation}
where $\eta_I$, $\nu_I$, and $\alpha_I$ are the critical exponents of the
Ising model.

In all dimensions, a simple argument 
based on the symmetry of the two-component cubic model~\cite{Korz-76}
shows that the cubic fixed point for $N=2$ has the same stability properties 
of the Ising one.
Indeed, for $N=2$, a $\pi/4$ internal rotation, i.e.
\begin{equation}
(\phi_1,\phi_2) 
\longrightarrow  {1\over \sqrt{2}} (\phi_1+\phi_2,\phi_1-\phi_2),
\label{sym1}
\end{equation}
maps the cubic Hamiltonian (\ref{Hphi4}) into a new one of the same form 
but with new couplings $(u_0',v_0')$ given by
\begin{equation}
u_0' = u_0+\case{3}{2}v_0, \qquad \qquad v_0' = -v_0.
\label{sym2}
\end{equation}
This symmetry maps the Ising fixed point onto the cubic one. 
So for all $d>2$, since the Ising point is unstable, the cubic point is 
unstable too, 
and the stable point is the isotropic one.
In two dimensions, this is no longer true. Indeed, one expects 
the cubic interaction to be truly marginal for $N=2$~\cite{jkkn-77,nr-82} and relevant for 
$N>2$ \cite{pn-76}, so $N_c = 2$ in two dimensions.

\subsection{The two dimensional case}
\label{2dcase}

As it is already clear from the previous subsection, in two-dimensions several 
new and interesting features appear. 
First of all we have no general argument to understand the stability
properties of the Ising 
fixed point, in fact the specific heat of the two-dimensional 
Ising model has a logarithmic divergence (i.e. $\alpha_I=0$).
The vanishing of this crossover exponents, and so the presence of a marginal 
operator, can bring us to several different scenarios. For example we may 
have that the Ising fixed point is marginally stable or unstable because of  
higher order corrections to RG equations, or that 
there exists a line of fixed points~(see Ref. \cite{Wegner-76} 
for a detailed review about the effects of marginal operators).

When the cubic anisotropy becomes marginal~(at $N=N_c=2$) \cite{pn-76}
the cubic fixed point is in the region 
with $v<0$~(cf. Eq. (\ref{sym2})),
and so it cannot coincide with the $O(N)$ one, 
as in the case of three and $4-\epsilon$ dimensions.

There are several studies on systems related to Landau-Ginzburg 
Hamiltonian (\ref{Hphi4}).
In Ref. \cite{jkkn-77} the model
\begin{equation}
{\cal H}= \sum_{\langle \mathbf{r}, \mathbf{r'}\rangle} 
J \, \mathbf{S} (\mathbf{r}) \mathbf{S} (\mathbf{r'}) +
\sum_{\mathbf{r}}  h_p \cos p \theta (\mathbf{r})
\label{h4}
\end{equation}
was considered,
where the first sum is only over nearest-neighbors and $\theta (\mathbf{r})$
is the angle that the two-component spin  $\mathbf{S} (\mathbf{r})$ forms with 
some arbitrary axis.
This system clearly belongs to the same universality class of $XY$ model with 
cubic anisotropy if $p=4$~(the ``field'' $h_4$ maps on $v_0$).

The phase diagram found in Ref. \cite{jkkn-77} consists of three 
distinct lines of fixed points showing
continuously varying exponents. The line with $h_4=0$ is the standard 
Kosterlitz-Thouless~(KT)  \cite{kt-73}. There are two lines with 
$h_4 \neq 0$,
starting from the $KT$ transition~(the end point of the KT line) and
continuing to infinitely, positive or negative, large values of $h_4$.
They map into each other with changing the sign of $h_4$.
The latter are lines of 
second-order phase transitions with conventional power law singularities 
characterized by $\eta=1/4$. The exponent $\nu$ diverges at the confluence
of these critical lines as
\begin{equation}
\nu\sim \frac{1}{|h_4|}
\end{equation}
when $h_4 \rightarrow 0$.
For this reason it is often said that the $XY$ model with cubic anisotropy 
has a non-universal behavior, although  there is the so-called 
weak-universality of Suzuki \cite{suzuki-74} (i.e. $\eta$ is constant).

In a successive work Kadanoff \cite{kadanoff-77} identifies the two 
fixed-point lines for $h_4 \neq 0$ with the dual line of the Ashkin-Teller~(AT)
and the eight-vertex Baxter model~(8V)~(see the Baxter book \cite{baxter-book} 
for a review about these models). One remark is necessary at this point:
the AT and the 8V models are equivalent on the dual line 
\cite{baxter-book} and they show a continuously varying exponent of the 
correlation length:
\begin{equation}
\nu=\frac{2-y}{3-2y},
\label{nuAT}
\end{equation}
where  $y$ is a parameter that appears in the Hamiltonians 
defined in the range $[0,4/3]$ for the AT \cite{ss-96} and  
$[0,2]$ for the 8V \cite{baxter-book}. The $O(2)$ multicritical point 
identified by Kadanoff is the F-model limit of the 8V that is characterized
by $y=3/2$, allowed to the 8V but  forbidden for the AT.

Another class of models related to the Landau-Ginzburg Hamiltonian 
(\ref{Hphi4}) is the discrete-cubic $N$-component model 
\cite{Aharony_77,Kim-etal_75,Kim-Levy_75,dr-79,riedel-81,nrs-83}. 
It is a short-range interacted 
system with $N$-component spins $\mathbf{S}_i$ pointing to the faces of an 
$N$-dimensional hypercube~(face-centered-cubic model).
The Hamiltonian may be written as:
\begin{equation}
{\cal H}= \sum_{\langle i,j \rangle} 
(C \delta_{a_i,a_j} s_i s_j+ P\delta_{a_i,a_j})
\label{dfcm}
\end{equation}
where $a_i$ is a Potts-like variable that determines which component of 
$\mathbf{S}_i$ is non-zero and $s_i$ an Ising variable that determines the 
sign of that component.
For $P=0$ the Hamiltonian (\ref{dfcm}) reduces to the $2N$-state Potts model, 
for $C=0$ to two decoupled $N$-state Potts model, and 
for $N=2$ to the AT model. 
The continuous cubic model (\ref{Hphi4}) reduces to 
the discrete one (\ref{dfcm}) in the limit of strong anisotropy
~($|v| \gg |u|$). In Ref. \cite{dr-79} it is shown that the iteration of RG
transformations enforces the continuous model, with $v_0<0$, to have spins  
pointing only 
to the faces of the hypercube, and so, also for finite anisotropy, the two 
models are equivalent at criticality.
The model (\ref{dfcm}) exhibits four competing possible types of critical 
behavior, related to the Ising model, the $N$-and $2N$-state Potts 
models and to a ``cubic'' fixed point.
In Refs. \cite{riedel-81,nrs-83} it was found that the critical behavior
of the discrete face-centered cubic model is $O(N)$-like for $N<N_c=2$,
AT-like for $N=2$, and characterized by a first-order phase transition for 
$N>2$. 
This result is not surprising since the model (\ref{dfcm}) is related to 
(\ref{Hphi4}) in the region with $v_0 <0$ where we expect a first-order
phase transition for $N>N_c=2$ and $O(N)$ behavior for $N<2$.

In the region with $v_0 >0$ the continuous model (\ref{Hphi4}) is related 
to the corner-cubic model in which
$N$-component spins $\mathbf{S}_i$  point to the corners of  an
$N$-dimensional hypercube. This model is equivalent to the 
face-centered one for $N=2$~(this is the symmetry (\ref{sym2}) for the
 Hamiltonian (\ref{Hphi4}) and the changing of the sign of $h_4$ for  
(\ref{h4})). For other values of $N$ the critical behavior of the 
face-centered- and the corner-cubic model is in principle very different.

This difference is clarified by the study of $N$-color Ashkin-Teller model,
first introduced by Grest and Widom \cite{gw-81}, that is equivalent to the 
corner-cubic model. The Hamiltonian of this model reads
\begin{equation}
{\cal H}=- \sum_{\langle i,j \rangle} 
\left[J\sum_{a=1}^N s_i^a s_j^a + J_4 \left[\sum_{a=1}^N s_i^a s_j^a\right]^2\right],
\label{ATN}
\end{equation}
where $s^a_i=\pm 1$ for $a=1,\dots ,N$ are $N$ Ising variables and the sum is
only over nearest-neighbor~(the sign of $J_4$ is the opposite of $u_0$ in Eq. 
(\ref{Hphi4})). 
In Ref. \cite{gw-81} it is concluded that the order $J_4^2$ in the RG equations
makes the decoupled Ising fixed point stable for perturbations with 
$J_4<0$~($u_0>0$) and unstable for $J_4>0$~($u_0<0$). In the latter case the
system flows away towards more negative values of $u$ and finally the 
transition is first order. This has been confirmed \cite{gw-81} by a 
Monte Carlo simulation for $N=3$. 
This argument was proved exactly in the 
$N \rightarrow \infty$ limit \cite{fradkin-84,bray-85}. 
After, using a mapping of the Hamiltonian (\ref{ATN}) onto the $O(N)$ 
Gross-Neveu model \cite{shankar-85} it was argued 
\cite{shankar-85,shalaev-89,shalaev-94,js-96}
that the critical behavior of the Ising fixed point is affected by 
logarithmic corrections.

Finally the three couplings $(N,2)$ model of Domany and Riedel 
\cite{dr-79}~(equivalent for $N=3$ to the $Z(6)$ model \cite{rwff-81}) 
reduces to the $N$-component discrete cubic model, for particular values of 
the parameters entering in the Hamiltonian. 
The model $(3,2)$ was studied in Refs. \cite{dr-79,rwff-81} with the 
Migdal-Kadanoff Renormalization Group approach~(MKRG). Unfortunately
the MKRG is expected to give neither precise values of the exponents nor the
correct nature of the transition, but only a good description of the phase
diagram.

\subsection{Order-disorder transition in adsorbed monolayers}
\label{sec2c}

We have already mentioned that the Hamiltonian (\ref{Hphi4}) describes some
order-disorder transitions in adsorbed monolayers. The great interest in 
these models is justified since they provide a unique possibility to study
experimentally a rich variety of two-dimensional systems.

Several experimental works have confirmed that these transitions belong
to the universality class of the Ising \cite{qpb-91,fc-93,mbrm-00}, 
the three- \cite{bretz-77,sp-94} and  
four-state Potts models \cite{pp-87,sp-94}, and
the $XY$ model with cubic anisotropy \cite{glb-90,bch-93,bkuoabe-85,footnote1},
according to their classification \cite{dswg-78,ds-79,class}.
Monte Carlo simulations and other numerical works confirm this 
scenario \cite{kkg-83,pdp-92,pdl-98}.

According to the standard classification, the critical behavior of 
the order-disorder transition of diatomic 
molecules on honeycomb lattice in the p$(2 \times 1)$ structure is described 
by the $N=3$ cubic model \cite{ds-79,dr-79}. 
The experimental investigation \cite{pp-89} of Oxygen on Ru$(001)$ 
shows that this transition has critical exponents in agreement with 
the three-state Potts model within about $10\% $.
Several numerical simulations~(after the first ones favoring a first order
phase transition \cite{mb-82,tc-89}) confirm this critical behavior 
\cite{cai-91,rj-93}.
So it may be possible that a fixed point of the Heisenberg model with 
cubic anisotropy has the critical behavior of the three-state
Potts model.

We want still to note the possibility for systems having exponents 
close to the four-state Potts model, to belong to the universality
class of the $XY$ model with cubic anisotropy and living near the 
four-state Potts transition characterized by $y=0$.

\section{THE FIXED DIMENSION PERTURBATIVE EXPANSION IN TWO DIMENSIONS}
\label{sec2}

\subsection{Renormalization of the theory}
\label{sec2a}

The fixed-dimension field-theoretical approach~\cite{Parisi-80} represents an 
effective procedure in the study of the critical properties 
of systems belonging to the $O(N)$ universality class (see, e.g., 
Ref.~\cite{ZJ-book}). The idea is to extend this procedure to models
where there are two $\phi^4$ couplings with different symmetry 
\cite{M-S-S-89,cpv-99}. One 
performs an expansion in powers of appropriately defined zero-momentum quartic
couplings and renormalizes the theory by a set of zero-momentum conditions
for the (one-particle irreducible) two-point and four-point correlation 
functions:

\begin{equation}
\Gamma^{(2)}_{ab}(p) = \delta_{ab} Z_\phi^{-1} \left[ m^2+p^2+O(p^4)\right],
\label{ren1}  
\end{equation}
\begin{eqnarray}
\Gamma^{(4)}_{abcd}(0) =& 
Z_\phi^{-2} m^2 \left[  
{u\over 3}\left(\delta_{ab}\delta_{cd} + \delta_{ac}\delta_{bd} + 
                \delta_{ad}\delta_{bc} \right)+\right. \nonumber \\
&\left. v \,\delta_{ab}\delta_{ac}\delta_{ad}\right].
\label{ren2}  
\end{eqnarray}
They relate the second-moment mass $m$, and the zero-momentum
quartic couplings $u$ and $v$ to the corresponding Hamiltonian parameters
$r$, $u_0$, and $v_0$:
\begin{equation}
u_0 = m^2 u Z_u Z_\phi^{-2},\qquad\qquad
v_0 = m^2 v Z_v Z_\phi^{-2}.
\end{equation}
In addition, one introduces the function $Z_t$ defined by the relation
\begin{equation}
\Gamma^{(1,2)}_{ab}(0) = \delta_{ab} Z_t^{-1},
\label{ren3}
\end{equation}
where $\Gamma^{(1,2)}$ is the (one-particle irreducible)
two-point function with an insertion of $\case{1}{2}\phi^2$.

From the pertubative expansion of the correlation functions
$\Gamma^{(2)}$, $\Gamma^{(4)}$, and $\Gamma^{(1,2)}$ and 
the above relations, one derives the functions $Z_\phi(u,v)$, 
$Z_u(u,v)$, $Z_v(u,v)$, and $Z_t(u,v)$ as a double expansion in $u$ and $v$.

The fixed points of the theory are given by 
the common  zeros of the $\beta$-functions
\begin{eqnarray}
\beta_u(u,v) &=& m \left. {\partial u\over \partial m}\right|_{u_0,v_0} ,
\nonumber \\
\beta_v(u,v) &=& m \left. {\partial v\over \partial m}\right|_{u_0,v_0} .
\label{bet}
\end{eqnarray}
The stability properties of the fixed points are controlled  by the 
eigenvalues $\omega_i$ of the matrix 
\begin{equation}
\Omega = \left(\matrix{\displaystyle \frac{\partial \beta_u(u,v)}{\partial u}
 &\displaystyle \frac{\partial \beta_u(u,v)}{\partial v}
\cr & 
 \cr \displaystyle \frac{\partial \beta_v(u,v)}{\partial u}
& \displaystyle  \frac{\partial \beta_v(u,v)}{\partial v}}\right)\; ,
\end{equation}
computed at the given fixed point:
a fixed point is stable if both eigenvalues are positive.
The eigenvalues $\omega_i$ are related to
the leading scaling corrections, which vanish as
$\xi^{-\omega_i}\sim |t|^{\Delta_i}$ where $\Delta_i=\nu\omega_i$.

One also introduces the functions
\begin{eqnarray}
\eta_\phi(u,v) &=& \left. {\partial \ln Z_\phi \over \partial \ln m}
         \right|_{u_0,v_0}
= \beta_u {\partial \ln Z_\phi \over \partial u} +
\beta_v {\partial \ln Z_\phi \over \partial v} , \label{eta1}\\
\eta_t(u,v) &=& \left. {\partial \ln Z_t \over \partial \ln m}
         \right|_{u_0,v_0}
= \beta_u {\partial \ln Z_t \over \partial u} +
\beta_v {\partial \ln Z_t \over \partial v}.\label{eta2}
\end{eqnarray}
Finally, the critical exponents are obtained from
\begin{eqnarray}
\eta &=& \eta_\phi(u^*,v^*),
\label{eta_fromtheseries} \\
\nu &=& \left[ 2 - \eta_\phi(u^*,v^*) + \eta_t(u^*,v^*)\right] ^{-1},
\label{nu_fromtheseries} \\
\gamma &=& \nu (2 - \eta),
\label{gamma_fromtheseries} 
\end{eqnarray}
where $(u^*,v^*)$ is the position of the stable fixed point.

\subsection{The four loop series}
\label{sec2b}

In this section we present the perturbative expansion of the 
RG functions (\ref{bet}), (\ref{eta1}), and (\ref{eta2}) up to four 
loops. The diagrams contributing to the two-point and four-point functions are
reported in Ref. \cite{nmb-77}. We do not calculate the integrals associated
to each diagram, but we use the numerical results compiled in Ref. 
\cite{nmb-77}. Summing all contributions with the right symmetry and 
group factors~(see Ref. \cite{cpv-99}) we obtain all RG functions.

The results are written in terms of the rescaled couplings
\begin{equation}
u \equiv  {8 \pi\over 3} \; R_N \; \bar{u},\qquad\qquad
v \equiv   {8 \pi\over 3} \; \bar{v} ,
\label{resc}
\end{equation}
where $R_N = 9/(8+N)$. We adopt this rescaling to have finite fixed point 
values in the limit $N\rightarrow \infty$.

The resulting series are
\begin{equation}
\bar{\beta}_{\bar{u}}  = -\bar{u} + \bar{u}^2 + {2\over 3} \bar{u} \bar{v}+ 
\bar{u} \sum_{i+j\geq 2} b^{(u)}_{ij} \bar{u}^i \bar{v}^j, 
 \label{bu}
\end{equation}
\begin{equation}
\bar{\beta}_{\bar{v}}  = -\bar{v} + \bar{v}^2 + {12\over 8+N} \bar{u} \bar{v}+
\bar{v} \sum_{i+j\geq 2} b^{(v)}_{ij} \bar{u}^i \bar{v}^j,  \label{bv} 
\end{equation}
\begin{equation}
\eta_\phi = \sum_{i+j\geq 2} e^{(\phi)}_{ij} \bar{u}^i \bar{v}^j,
\label{etaphi}
\end{equation}
\begin{equation}
\eta_t = -{2 (2+N)\over (8+N) } \bar{u} -{2\over 3} \bar{v} +
\sum_{i+j\geq 2} e^{(t)}_{ij} \bar{u}^i \bar{v}^j
\label{etat}
\end{equation}
where
\begin{equation}
\bar{\beta}_{\bar{u}}= {3\over 16\pi} \, R_{2N}^{-1}\beta_{u},\qquad\qquad
 \bar{\beta}_{\bar{v}}= {3\over 16\pi} \, R_{2N}^{-1}\beta_{v}.
\label{resc2}
\end{equation}
The coefficients  $b^{(u)}_{ij}$,  $b^{(v)}_{ij}$, $e^{(\phi)}_{ij}$, and 
$e^{(t)}_{ij}$  are reported in the Tables \ref{t1}, \ref{t2}, \ref{t3}, and 
\ref{t4}.
Note that due to the  rescaling (\ref{resc2}), the matrix element of  $\Omega$ 
 are two times the derivative of $\bar{\beta}$ with respect to $\bar{u}$ 
and $\bar{v}$.

We have verified the exactness of our series by the following relations:

(i) $\bar{\beta}_{\bar{u}}(\bar{u},0)$, $\eta_\phi(\bar{u},0)$ 
and $\eta_t(\bar{u},0)$ reproduce
the corresponding functions of the O($N$)-symmetric
model~\cite{B-N-G-M-77,os-00};

(ii) $\bar{\beta}_{\bar{v}}(0,\bar{v})$, $\eta_\phi(0,\bar{v})$ and 
$\eta_t(0,\bar{v})$ reproduce
the corresponding functions of the Ising-like ($N=1$) $\phi^4$ theory;

(iii) The following relations hold for $N=1$: 
\begin{eqnarray}
&& \bar{\beta}_{\bar{u}}(u,x-u) + \bar{\beta}_{\bar{v}}(u,x-u) =\bar{\beta}_{\bar{v}}(0,x),\\
&&\eta_\phi(u,x-u) = \eta_\phi(0,x),\nonumber\\
&&\eta_t(u,x-u) = \eta_t(0,x).\nonumber
\end{eqnarray}

(iv) For $N=2$, using the symmetry 
(\ref{sym1}) and (\ref{sym2}), and 
taking into account the rescalings (\ref{resc}), one can easily obtain
the identities 
\begin{eqnarray}
&& \bar{\beta}_{\bar{u}}(\bar{u}+\case{5}{3}\bar{v},-\bar{v}) + 
{5\over 3}\bar{\beta}_{\bar{v}}(\bar{u}+\case{5}{3}\bar{v},-\bar{v}) = 
\bar{\beta}_{\bar{u}}(\bar{u},\bar{v}),\\
&& \bar{\beta}_{\bar{v}}(\bar{u}+\case{5}{3}\bar{v},-\bar{v}) = -\bar{\beta}_{\bar{v}}(\bar{u},\bar{v}), \nonumber\\
&&\eta_\phi(\bar{u}+\case{5}{3}\bar{v},-\bar{v}) = \eta_\phi(\bar{u},\bar{v}),\nonumber\\
&&\eta_t(\bar{u}+\case{5}{3}\bar{v},-\bar{v}) = \eta_t(\bar{u},\bar{v}).\nonumber
\end{eqnarray}
These relations are exactly satisfied by our four-loop series.
Note that, since the Ising fixed point is $(0,g^*_I)$, and $g^*_I$ is known 
with very high precision from Ref. \cite{chpv-00}
\begin{equation}
g^*_I=1.7543637(25), \label{gI*}
\end{equation}
the above symmetry gives us 
the location of the cubic fixed point: $(\case{5}{3} g^*_I,-g^*_I)$.

(v) In the large-$N$ limit 
the critical exponents of the cubic fixed point 
are related to those of the Ising model:
$\eta=\eta_I$ and $\nu = \nu_I$.
One can easily see
that, for $N\to\infty$,
$\eta_\phi(u,v) = \eta_I(v)$, where $\eta_I(v)$ is 
the perturbative series that determines the exponent $\eta$ of the Ising model.
Therefore, the first relation is trivially true. On the other hand,
the second relation $\nu = \nu_I$ is not identically
satisfied by the series, and is verified only at the critical 
point \cite{cpv-99}.

(vi) For $N=0$ the series reproduce the results  of Ref. \cite{mayer-89}.

\begin{table}[tbp]
\squeezetable
\caption{
The coefficients $b^{(u)}_{ij}$, cf. Eq. (\ref{bu}).
}
\label{t1}
\begin{tabular}{cl}
\multicolumn{1}{c}{$i,j$}&
\multicolumn{1}{c}{$(N+8)^i b^{(u)}_{ij}$}\\
\tableline \hline
2,0 &$-(10.3350 +47.675 N)$\\
1,1 &$-8.39029$  \\
0,2 &$-0.21608$ \\ \hline
3,0 &$524.377 + 149.152 N + 5.00028 N^2 $\\
2,1 &$144.813 +7.27755 N$       \\  
1,2 &$10.0109 +0.0583278 N $     \\
0,3 &$0.231566             $    \\
\hline
4,0 &$-( 7591.108 + 2611.15 N + 179.697 N^2 + 0.0888427 N^3) $ \\
3,1 &$-( 2872.09 + 291.255 N - 0.126814 N^2               ) $ \\
2,2 &$-( 330.599 + 5.97086 N                              ) $ \\
1,3 &$-( 16.0559 + 0.0578955 N                            ) $ \\
4,0 &$- 0.311695                                          $ 
\end{tabular}
\end{table}

\begin{table}[tbp]
\squeezetable
\caption{The coefficients $b^{(v)}_{ij}$, cf. Eq. (\ref{bv}).
}
\label{t2}
\begin{tabular}{cl}
\multicolumn{1}{c}{$i,j$}&
\multicolumn{1}{c}{$(N+8)^i b^{(v)}_{ij}$}\\
\tableline \hline
2,0 &$-(92.6834+5.83417 N)$\\
1,1 &$-17.392$  \\
0,2 &$-0.716174$ \\ 
\hline
3,0 & $ 1228.63 + 118.504 N - 1.83156 N^2 $ \\
2,1 & $ 358.882 + 2.84758 N               $ \\
1,2 & $ 31.4235                           $ \\
0,3 & $ 0.930766                          $ \\
\hline
4,0 & $ -(20723.1 + 2692.0 N + 25.4854 N^2 + 0.824655 N^3) $ \\
3,1 & $ -(8273.28 + 233.78 N - 0.574757 N^2) $ \\
2,2 & $ -(1134.8 + 1.91402 N       )         $ \\
1,3 & $ -68.4022                           $ \\
0,4 & $ -1.58239                           $ 
\end{tabular}
\end{table}

\begin{table}[tbp]
\squeezetable
\caption{
The coefficients $e^{(\phi)}_{ij}$, cf. Eq.(\ref{etaphi}).
}
\label{t3}
\begin{tabular}{cl}
\multicolumn{1}{c}{$i,j$}&
\multicolumn{1}{c}{$(N+8)^i e^{(\phi)}_{ij}$}\\
\tableline \hline
2,0 &$0.917086 \,(N+2)$\\
1,1 &$0.611391$  \\
0,2 &$0.033966$ \\ 
\hline
3,0  & $ -0.054609 \,(N+2) (N+8) $\\
2,1  & $ -0.054609 \,(N+8)               $\\
1,2  & $ -0.054609                     $\\
0,3  & $ -0.002023                    $\\
\hline
4,0  & $ (N+2)(26.7676+ 4.24178 N - 0.092684 N^2) $\\
3,1  & $ 27.6901 + 5.6557 N - 0.123579 N^2               $ \\
2,2  & $ 5.404243 + 0.132800 N         $ \\
1,3  & $ 0.410151                      $ \\
0,4  & $ 0.011393                      $ 
\end{tabular}
\end{table}

\begin{table}[tbp]
\squeezetable
\caption{
The coefficients $e^{(t)}_{ij}$, cf. Eq. (\ref{etat}).
}
\label{t4}
\begin{tabular}{cl}
\multicolumn{1}{c}{$i,j$}&
\multicolumn{1}{c}{$(N+8)^i e^{(t)}_{ij}$}\\
\tableline \hline
2,0 &$6.751258 \,(N+2)$\\
1,1 &$4.500839$  \\
0,2 &$0.250047$ \\ 
\hline
\hline
3,0  & $ -(96.7105 + 65.1686 N + 8.40668 N^2) $ \\
2,1  & $ -(48.3553 + 8.40668 N)               $ \\
1,2  & $ -(6.19023 + 0.116656 N)              $ \\
0,3  & $ -0.233588                          $ \\
\hline
4,0  & $ 1135.05 + 844.5 N + 139.656 N^2 + 0.583377 N^3 $ \\
3,1  & $ 756.697 + 184.652 N + 0.777836 N^2 $ \\
2,2  & $ 149.467584 + 7.553459 N            $ \\
1,3  & $ 11.515397 + 0.115791 N $ \\
0,4  & $ 0.323089               $ 
\end{tabular}
\end{table}

\subsection{Resummations of the series}

The field theoretic perturbative 
expansion generates asymptotic series that must be resummed to extract the 
physical information about the critical behavior of the real systems.

Exploiting the property that these series are Borel summable for $\phi^4$ 
theories in two and three dimensions  \cite{bores}, one can resum these 
perturbative expressions considering the Borel transform combined with a 
method for its analytical extension. 
In the case of the $O(N)$ symmetric model with only a coupling $g$, all
perturbative series are of the form $F(g)=\sum f_kg^k$. Exploiting 
the knowledge of the large order behavior of the coefficients 
$f_k$~(Ref. \cite{ZJ-book})
\begin{equation}
f_k \sim k! \,(-a)^{k}\, k^b \,\left[ 1 + O(k^{-1})\right] \qquad
{\rm with}\qquad a = - 1/g_b,
\label{lobh}
\end{equation}
(a large order behavior related to the singularity $g_b$ of the Borel transform
 closest to the origin) one can perform the following mapping~\cite{L-Z-77}
\begin{equation}
y(g) = {\sqrt{1 - g/g_b} - 1\over \sqrt{1 - g/g_b} + 1 }
\end{equation}
to extend the Borel transform of $F(g)$ to all positive values of $g$ . 
The singularity $g_b$ depends only on the 
considered model and can be obtained from a steepest-descent calculation in which the relevant saddle point is a finite-energy solution (instanton) of the classical field equations with negative coupling~\cite{Lipatov-77,B-L-Z-77}. Instead the coefficient $b$ depends on which Green's function is considered.

Note that the function $F(g)$ can be Borel summable only if there are no
singularities of the Borel transform on the positive real axis.

This resummation procedure has worked successfully for the $O(N)$ symmetric 
theory, for which accurate estimates for the critical exponents and other 
physical quantities have been obtained \cite{L-Z-77,L-Z-87,G-Z-98}. 

For this reason we want to extend the resummation procedure cited above to
multicoupling models, as it has been done for the three-dimensional cubic model 
\cite{cpv-99,os-00} and for the frustrated system with non-collinear order in 
two dimensions \cite{cp-01}.
Considering a double expansion in $\bar{u}$ and $\bar{v}$ at fixed 
$z=\bar{v}/\bar{u}$, and  studying the large order behavior (following the 
same procedure used in Refs. \cite{cpv-99,cp-01}) of the new expansion 
in powers of $\bar{u}$ to 
calculate the singularity of the Borel transform closest to the origin 
$\bar{u}_b$ we have
\begin{eqnarray}
{1\over \bar{u}_b} &= - a \left( R_N  + z \right)
\qquad & {\rm for} \qquad  0< z ,
\label{bsing} \\
{1\over \bar{u}_b} &= - a \left( R_N  + {1\over N} z \right)
\qquad & {\rm for} \qquad  0 > z > - {2 N R_N\over N+1},
\nonumber
\end{eqnarray}
where $a = 0.238659217\dots \, \,$.

Note that the series in powers of $\bar{u}$ keeping $z$ fixed
is not Borel summable for $\bar{u}>0$ and  $z< -R_N$.
This fact will not be a real limitation for us,
since we will only consider values of $z$ such that $\bar u_b<0$.
It should be noted that these results do not apply to the case $N=0$. 
Indeed, in this case, additional singularities in the 
Borel transform are expected \cite{nrN0}.

The exponent $b$ in Eq. (\ref{lobh}) is related to the number of 
symmetries broken by the classical solution~\cite{B-L-Z-77}.
It depends on the quantity considered. In the cubic model, for $v\not=0$, 
we have $b=2$ for the function $\eta_\phi$, and 
$b=3$ for the $\beta$-functions and $\eta_t$. 
For $v=0$, we recover the results of the O($N$)-symmetric model, that is 
$b=(3 + N)/2$ for $\eta_\phi$, and $b=(5+N)/2$ for 
the $\beta$-function and $\eta_t$~\cite{L-Z-77}. 

Good estimates of the critical exponent could be obtained 
also using a Pad\'e Borel analysis of the series, as shown in the case of  
the two-dimensional $O(N)$ models \cite{B-N-G-M-77,os-00} 
and the random Ising \cite{mayer-89}.

An important issue in the fixed dimension approach to critical phenomena~(and
in general of all the field theoretical methods) 
concerns the analytic properties of the $\beta$-functions. 
As shown in Ref. \cite{cccpv-00} for the $O(N)$ model, the presence of 
confluent singularities in the zero of the perturbative $\beta$ function
causes a 
slow convergence of the resummation of the perturbative series to the correct
fixed point value. The $O(N)$ two-dimensional field-theory estimates of 
physical quantities 
\cite{L-Z-77,os-00} are less accurate than  the three-dimensional 
ones, due to the stronger nonanalyticities at the fixed point
\cite{cccpv-00,Nickel-82,Nickel-91,pv-98}.
In Ref. \cite{cccpv-00} it is shown that the nonanalytic terms may cause
large imprecisions in the estimate of the exponent related to the leading 
correction to the scaling $\omega$; instead the result for the 
fixed point value is a rather
good approximation of the correct one~(if one compare the field theoretical
results for the four point renormalized coupling in the $g$-expansion 
\cite{os-00} and in the $\epsilon$-expansion \cite{pv-98,pv-00} with
the estimates of other non-perturbative methods 
\cite{chpv-00,balogetal,kim-95}, 
one finds that the systematic error is always less 
then 10\%). 
We think that this scenario holds also for the cubic models.

\section{FOUR-LOOP EXPANSION ANALYSIS}
\label{sec3}

\subsection{The analysis method}

In order to study the critical properties of the continuous anisotropic cubic
model ({\ref{Hphi4}), we use two different resummation procedures: 
the Pad\'e-Borel method
and the conformal mapping of the Borel-transformed series.

Explicitly, let us consider an $l$-loop series in $\ub$ and $\vb$ of the form
\begin{equation}
R(\bar{u},\bar{v}) =\, \sum_{k=0}^l \sum_{h=0}^{l-k} R_{hk} \ub^h \vb^k=\, 
\sum_{k=0}^l R_{k}(z) \ub^k,
\end{equation}
where $R$ is one of the RG functions and 
$z$ is the ratio $\vb / \ub$ that we will consider always fixed to same 
value. In this manner we have an asymptotic series of only one variable $\ub$,
depending on the additional parameter $z$ that have to run from $0$ to 
$\infty$ in order to reproduce all the quadrant with $\ub,\vb \geq 0$.
In order to use a variable defined in a finite range we set 
\begin{equation}
z=\tan \frac{\pi}{2} x,
\label{zversusk}
\end{equation}
with $x \in [0,1]$. In the follow we fix $x=k/N_{max}$, where $k$ is an integer 
number running from 0 to $N_{max}$ 
~(we will fix $N_{max}=25$).

In the Pad\'e-Borel method we consider the Pad\'e approximants $[L/M]$ of the
Borel-Leroy transform of $R(\ub ,z)$. 
Explicitly the approximants of $R$ are 
\begin{eqnarray}
E({R})(L,M,b;\bar{u},z)= 
    \int_0^\infty dt\,t^b e^{-t} \frac{N_L (\ub t,z)}{D_M (\ub t,z)}
\label{padeapprox}
\end{eqnarray}
where $N_L$ and $D_M$ are two polynomials of degree $L$ and $M$ respectively, 
with $L+M \leq l$, 
that are determined by the condition that the expansion of 
$E({R})(L,M,b;\bar{u},z)$ in powers of $\bar{u}$ gives $R(\bar{u},z)$ to 
order $L+M$.

In this manner we have several approximants of the function $R(\ub,\vb)$ 
with varying the three parameters $b, \, L$, and $M$.
As usual in this case\cite{os-00} the best estimates of the resummed function 
are given by the diagonal and near-diagonal approximants~(i.e. for our 
four-loop series we have three reasonable choices $[2,2]$, $[3,1]$, 
and $[2,1]$).
Then we search the value of $b$ (called $b_{opt}$) integer or half-integer 
minimizing the differences between the considered approximants; a reasonable
estimate will be the mean value of $E(b_{opt},L,M)$ on all the values of
$L$ and $M$ considered. Then we could take as error bar the deviations from
the mean value of all the approximants with $b_{opt}-1 \leq b \leq b_{opt}+1$
(always considering integer and half-integer values of $b$).

In the analysis using the conformal-mapping method we essentially follow the 
procedure used in Ref. \cite{cpv-99}. We exploit the knowledge of the value 
of the singularity of the Borel transform closest to the origin (a value given 
in the previous section), and we generate a set of approximants to our 
asymptotic series, varying the two parameters $\alpha$ and $b$ appearing in 

\begin{eqnarray}
E({R})_p(\alpha,b;\ub,z)&=& \sum_{k=0}^l 
  B_k(\alpha,b;z) \nonumber \\
 & \times &  \int_0^\infty dt\,t^b e^{-t} 
  {y(\ub t;z)^k\over [1 - y(\ub t;z)]^\alpha},
\label{approx}
\end{eqnarray}
where
\begin{equation}
y(x;z) = {\sqrt{1 - x/\overline{u}_b(z)} - 1\over 
          \sqrt{1 - x/\overline{u}_b(z)} + 1}.
\end{equation}
The coefficients $B_k$ are determined by the condition that the expansion of 
$E({R})_p(\alpha,b;\ub,z)$ in powers of $\bar{u}$ and $\bar{v}$ 
gives $R(\bar{u},\bar{v})$ to order $l$.

The procedure to choose the range of the parameters $\alpha$ and $b$  used to
find good estimates and reasonable error bars of the various quantities is the
same of Ref. \cite{cpv-99}.

\subsection{Stability properties of the $O(N)$ and the Ising fixed points}

First of all, we  analyze the stability properties of the $O(N)$-symmetric
fixed point. 
Since 
\begin{equation}
\frac{\partial\beta_\vb}{\partial \ub}(\ub,0) = 0, 
\end{equation}
the eigenvalues are simply
\begin{equation}
\omega_1 =2\, {\partial\bar{\beta}_\ub\over \partial \ub}(\ub^*,0),\qquad\qquad
\omega_2 =2\, {\partial\bar{\beta}_\vb\over \partial \vb}(\ub^*,0),
\end{equation}
where $\ub^*$ is the fixed-point value of the $O(N)$ vector model
of which accurate estimates are available from the five-loop analysis of
Ref. \cite{os-00}. 
The exponent $\omega_1$ is the usual exponent of the 
$O(N)$-symmetric theory \cite{ZJ-book} that gives the first correction to the 
scaling. For all $N > 2$ it is known that $\omega_1$  assumes the constant
value $2$~(see Ref. \cite{cccpv-00} and references therein). 
Instead $\omega_2$ is the eigenvalue
that determines the stability of the fixed point respect to an anisotropic 
cubic perturbation.

In Table \ref{omon} we report the results for $\omega_2$ for several values of
$N$.
It is quite evident that the $O(N)$ fixed 
point is unstable for $N \geq 3$~(we note that for small $N$ these values
are very close to zero). 
In the limit $N\rightarrow \infty$ it holds $\omega_2=-2$.
For $N=2$ our result is compatible with the presence of a marginal 
operator, i.e., $\omega_2=0$.

\begin{table}[t]
\caption{Half of the exponent $\omega_2$ at the $O(N)$ fixed point. 
CM is the value 
obtained using conformal mapping technique and PB the one using Pad\'e-Borel.}
\begin{tabular}{r|c|cc}
N & $u^*_{O(N)}$ & CM &  PB \\
\tableline \hline
2 & 1.80(3)&$ 0.03(3)$  & $ 0.06(4)$  \\
3 & 1.75(2)&$-0.08(3)$  & $-0.07(3)$ \\
4 & 1.70(2)&$-0.18(4)$  & $-0.17(5)$ \\
8 & 1.52(1)&$-0.45(5)$  & $-0.44(6)$   
\end{tabular}
\label{omon}
\end{table}

Then we focus our attention on  the stability properties of the Ising
fixed point. 
Also in this case the eigenvalues are simply:
\begin{equation}
\omega_1 =2\, {\partial\bar{\beta}_\vb\over \partial \vb}(0,\vb^*),\qquad\qquad
\omega_2 =2\, {\partial\bar{\beta}_\ub\over \partial \ub}(0,\vb^*),
\end{equation}
where $\vb^*$ is the fixed-point value of the Ising model (\ref{gI*}).
The exponent $\omega_1$ gives the correction to the scaling of the Ising
model and it should be equal to $7/4$ \cite{cccpv-00}, while $\omega_2$ is 
the eigenvalue determining the stability of this fixed point.
We find that the series $\omega_2(\vb)$ is independent from $N$~(we
already know that its fixed point value must be equal to $\alpha_I$ for all
$N$):
\begin{equation}
{\omega_2^I(\vb)\over 2}=
-1+ {2\over3} \vb  - 0.2161 \vb^2  + 0.23157 \vb^3  - 0.31169 \vb^4 \, .
\end{equation}

The fixed point value of this exponent is $\omega_2^I/2=-0.10(5)$,
using the conformal mapping method, and $-0.08(5)$, using the Pad\'e-Borel 
analysis. These values are quite close to $-\alpha_I=0$;
we attribute this small discrepancy to the presence of  nonanalyticities 
at the zero of the $\bar{\beta}_\vb$. To support this thesis we note that 
the value
of $\omega_2/2$ is approximately the same of the resummation of the exponent 
$-\alpha_I=-0.081$ found in Ref. \cite{os-00}.

\subsection{Evaluation of non-trivial fixed points}

Since the Ising fixed point has always a marginal operator and the $O(N)$ is 
unstable for $N\geq 3$,
we search for the presence of non-trivial fixed points in the region 
$\ub,\vb >0$.

\begin{figure}[t]
\centerline{\psfig{height=6truecm,width=8.6truecm,file=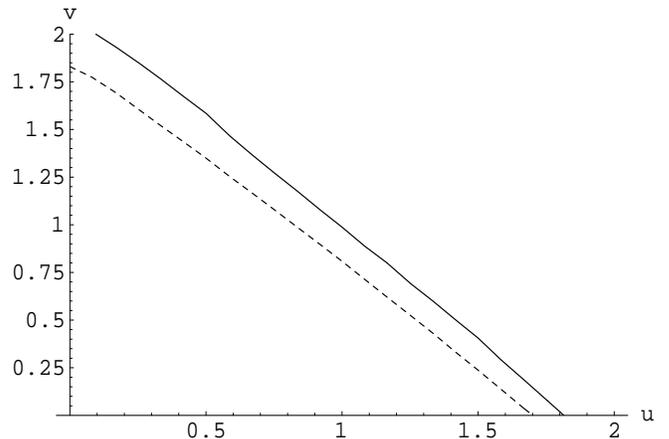}}
\caption{Zeros of the $\bar{\beta}$-functions for $N=2$ in the $(\bar{u},\bar{v})$ plane using the conformal mapping method.
The straight line correspond to zeros of $\bar{\beta}_{\bar{u}}(\bar{u},\bar{v})$ and the dashed one to the zeros of $\bar{\beta}_{\bar{v}}(\bar{u},\bar{v})$.}
\label{cm2}
\end{figure}

\begin{figure}[b]
\centerline{\psfig{height=6truecm,width=8.6truecm,file=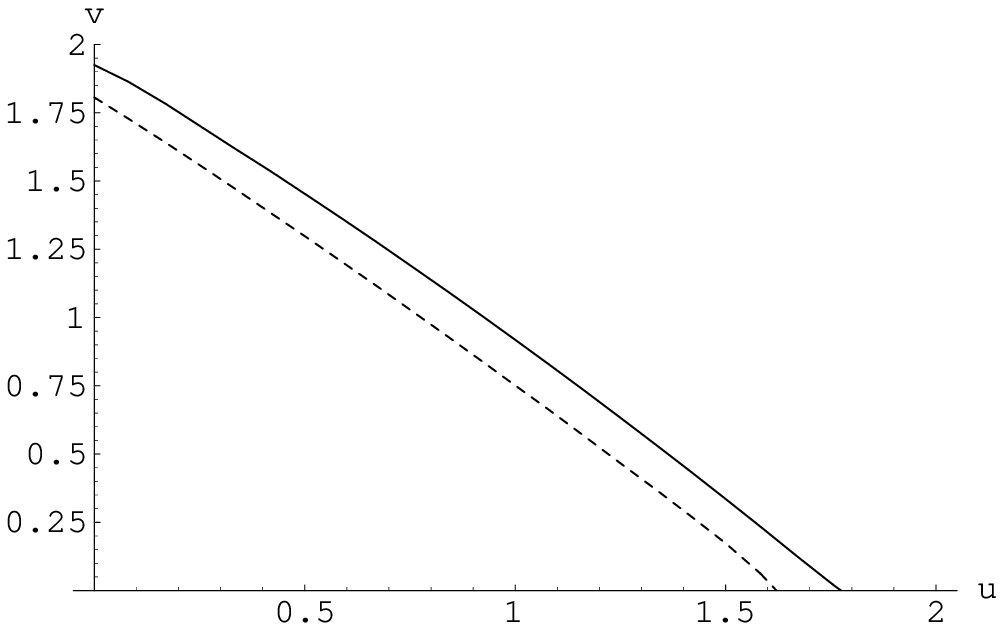}}
\caption{Zeros of the $\bar{\beta}$-functions for $N=2$ in the $(\bar{u},\bar{v})$ plane using the Pade-Borel analysis.
The straight line correspond to zeros of $\bar{\beta}_{\bar{u}}(\bar{u},\bar{v})$ and the dashed one to the zeros of $\bar{\beta}_{\bar{v}}(\bar{u},\bar{v})$.}
\label{pb2}
\end{figure}

In Figs. \ref{cm2} and \ref{pb2} we show the zeros of the $\beta$ functions 
for N=2  using the conformal mapping technique and the Pad\'e-Borel analysis 
respectively.
The curves of zeros of $\bar{\beta}_{\ub}$ and $\bar{\beta}_{\vb}$ seem 
parallel and do not have any intersection~(in these figures, and in all the
others, the error bar are not shown to make them more readable).
The distance between the two
lines is always less than the error bar~(that is of the order of $0.1$ for
each line). So our result is compatible with the presence of the line
of fixed points conjectured by Kadanoff \cite{kadanoff-77}, joining the
decoupled Ising and the $O(2)$ fixed points. 
In the Kadanoff identification this line of fixed points is the dual line of
the 8V model in the region that goes from the F-model limit~(O(2) multicritical
point with $y=3/2$) to the decoupled Ising point~($y=1$).
This line of fixed points is identifiable with the one of Ref. \cite{jkkn-77}
with $h_4<0$.

\begin{figure}[t]
\centerline{\psfig{height=6truecm,width=8.6truecm,file=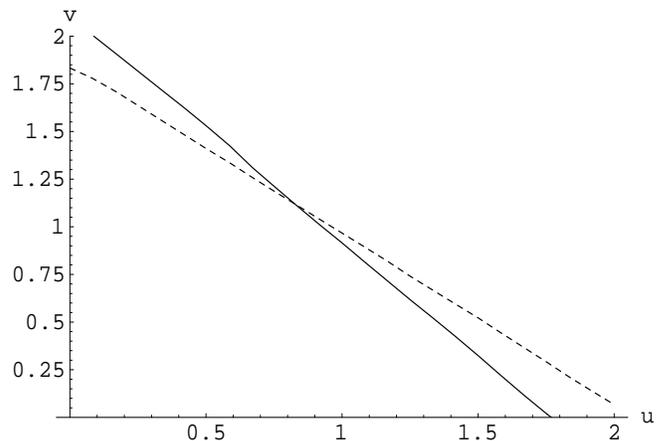}}
\caption{Zeros of the $\bar{\beta}$-functions for $N=3$ in the $(\bar{u},\bar{v})$ plane using the conformal mapping method.
The straight line correspond to zeros of $\bar{\beta}_{\bar{u}}(\bar{u},\bar{v})$ and the dashed one to the zeros of $\bar{\beta}_{\bar{v}}(\bar{u},\bar{v})$.}
\label{cm3}
\end{figure}

For the symmetry (\ref{sym2}) the fixed-point line continues for negative 
$\vb$ and reaches the ``cubic'' fixed point~(that in this case is a standard 
Ising) located at $(5/3 \, g^*_I,-g^*_I)$. This second line is the 
one of Ref. \cite{jkkn-77} with $h_4>0$ that is isomorphic to the other one 
with $h_4<0$.
We do not know if the line continues in the region with negative $\ub$, that 
may be a region of second order phase transition, having the critical 
exponent of the 8V model with $y<1$~(only for $v>-u$, from the  stability 
condition).
We tried to perform the resummation also for these value of $\ub,\vb$ but we 
obtained strongly oscillating results.

For the other values of $N$ we find one fixed 
point at the intersection of the two curves of zeros of the $\beta$-functions,
that is
the usual cubic fixed point, analytic continuation of the one found in the 
$\epsilon$-expansion. For example the results of the conformal mapping
analysis for $N=3$ and of the Pad\'e-Borel for $N=4$ are shown in Figs. 
\ref{cm3} and \ref{pb4} respectively. All our results about 
the cubic fixed points are summarize in Table \ref{tfp}. We note that,
within the error bar, there is full agreement between the two methods 
\cite{footnote3}.

\begin{figure}[b]
\centerline{\psfig{height=6truecm,width=8.6truecm,file=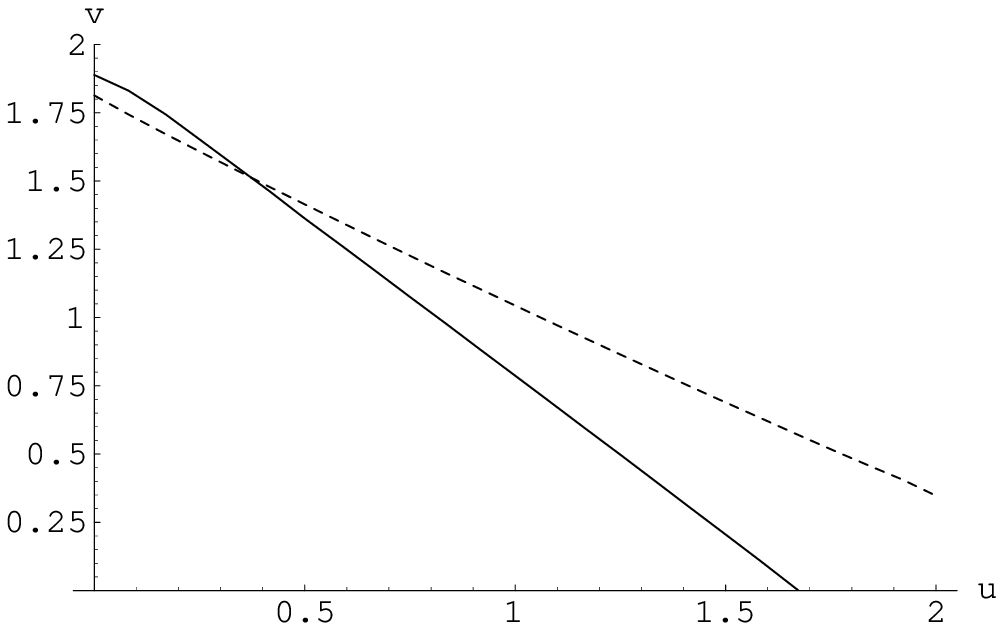}}
\caption{Zeros of the $\bar{\beta}$-functions for $N=4$ in the $(\bar{u},\bar{v})$ plane using the Pade-Borel analysis.
The straight line correspond to zeros of $\bar{\beta}_{\ub}(\ub,\vb)$ and the dashed one to the zeros of $\bar{\beta}_{\vb}(\ub,\vb)$.}
\label{pb4}
\end{figure}

\begin{table}[t]
\caption{Cubic fixed points  for $N \geq 3$.}
\label{tfp}
\begin{tabular}{r|c|c}
N & CM &  PB\\
  &$(\bar{u}^*,\bar{v}^*)$& $(\bar{u}^*,\bar{v}^*)$\\
\tableline \hline
8 & (0.24(8),1.72(10))  &(0.15(9),1.72(12)) \\
4 & (0.54(10),1.43(8))  &(0.38(11),1.51(9)) \\
3 & (0.83(12),1.12(9))  &(0.71(12),1.17(9))
\end{tabular}
\label{tabexp}
\end{table}

In the limit $N\rightarrow \infty$ the series $\bar{\beta}_{\ub}(\ub,\vb)$
and $\bar{\beta}_{\vb}(\ub,\vb)$ simplify to
\begin{eqnarray}
\bar{\beta}_{\vb}(\ub,\vb)\rightarrow& \bar{\beta}_{\mbox{Ising}}(\vb),\\
\bar{\beta}_{\ub}(\ub,\vb)\rightarrow& \ub [P_1(\vb)-\ub P_2(\vb)],
\end{eqnarray}
with
\begin{equation}
P_1(\vb)=-1+\frac{2}{3} \vb -0.21608 \vb^2+0.23157 \vb^3-0.31168 \vb^4
\end{equation}
and
\begin{equation}
P_2(\vb)= 1+0.058328 \vb^2-0.057896 \vb^3 \, . 
\end{equation}

The zero of $\bar{\beta}_{\vb}(\ub,\vb)$ is constant with varying $\ub$ at
the value $\vb^*= \bar{g}_I^*$ (\ref{gI*}), and so 
the $\ub$ coordinate of the cubic fixed point is simply:
\begin{equation}
u^*_{N\rightarrow \infty}=\frac{P_1(\bar{g}_I^*)}{P_2(\bar{g}_I^*)}=0.09(4).
\label{uinf}
\end{equation}
We note that the small value of $u^*$ for $N\rightarrow \infty$ 
does not exclude that the Ising and the cubic fixed points coalesce in
this limit.

We want to add that in Ref. \cite{shalaev-94}, using  an argument based on 
conformal field theory, it is conjectured that the cubic fixed point 
and the Ising one merge for arbitrary N. 
Our analysis seems to contradict this statement.

\subsection{Stability properties of the cubic fixed point}

\begin{figure}[b]
\centerline{\psfig{height=6truecm,width=8.6truecm,file=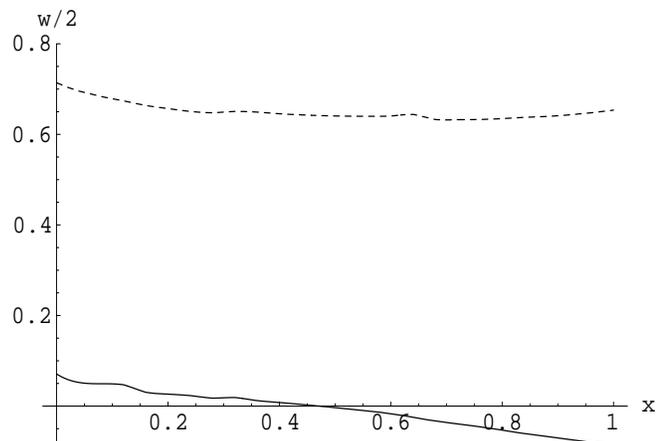}}
\caption{Values of $\omega_1/2$~(dashed line) and $\omega_2/2$~(straight line)
with varying the parameter $x$ (\ref{zversusk}). }
\label{omn2}
\end{figure}

For $N=2$ we check that one eigenvalue of the stability matrix $\Omega$  
vanish on all the line of fixed points. From Figs. \ref{cm2}, \ref{pb2} it is 
evident that 
this line is almost straight, so for our calculation we evaluate the 
$\Omega$ matrix on the line joining the Ising fixed point~(at $(0,g^*_I)$) 
and the $O(2)$~(at $(1.80(3),0)$ from Ref. \cite{os-00}).
Within the precision of our calculation we confirm that on this line 
one marginal operator exists~(see Fig. \ref{omn2}, the uncertainty of 
$\omega_2/2$ is constantly of the order of $0.05$).

\begin{figure}[b]
\centerline{\psfig{width=8.6truecm,file=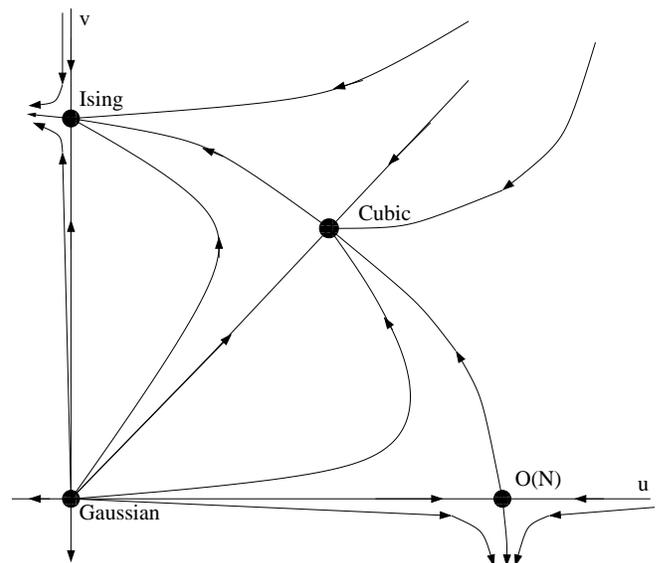}}
\caption{
Conjectured renormalization-group flow in the coupling plane $(u,v)$ for $N >2 $.}
\label{rgflow2}
\end{figure}

The results for the eigenvalues of the $\Omega$ matrix at the cubic fixed point
for $N\geq 3$ are summarize in Table \ref{omn>2}. 
The eigenvalue $\omega_2$ is positive for all considered values of $N$, but
it is very close to zero. Being the Ising fixed point stable against 
perturbations with positive $u$, it is not possible that the cubic fixed point
is stable for ``left'' perturbations. We believe that our results are 
compatible with the presence of a marginal operator also at the cubic fixed 
point and that the very small discrepancy of $\omega_2$ with respect to zero 
is due to nonanalytic terms.

We conjecture that the RG flow in the coupling plane $u,v$ for $N \geq 3$ is 
the one sketched in Fig. \ref{rgflow2}. 
For $u,v<0$ we expect a first order phase transition.
For $0<v/u \leq v^*/u^*$ the transition is second order and its critical 
behavior is characterized by the cubic fixed point. 
For $v/u \geq v^*/u^*$ the transition is
governed by the Ising fixed point, i.e., with the exponents of the Ising
model more logarithmic corrections. 

\begin{table}[t]
\caption{The half of eigenvalues of the $\Omega$ matrix for $N \geq 3$ at the 
cubic fixed point. }
\begin{tabular}{r|cc|c}
N       &       CM            &         PB            &  pseudo-$\epsilon$ \\
 &($\omega_1/2,\omega_2/2$)&($\omega_1/2,\omega_2/2$)&($\omega_1/2,\omega_2/2$)\\
\tableline \hline
$\infty$&                     & $(0.64(10),0.11(8))$   & $(0.65(3),0.07(4))$\\
8       &                     & $(0.70(10),0.07(6))$  & $(0.72(3),0.08(4))$ \\
4       & $(0.65(15),0.11(6))$& $(0.70(15),0.06(6))$  & $(0.79(4),0.06(4))$ \\
3       & $(0.70(7),0.09(6))$ & $(0.70(10),0.04(5))$  & $(0.75(6),0.05(5))$
\end{tabular}
\label{omn>2}
\end{table}

To support this scenario we also consider the pseudo-$\epsilon$ expansion
using substantially the same procedure of Ref. \cite{fhy-00}. By using  
this trick
we have a smallest error bar since we avoid the uncertainty of the fixed 
point~(for a detailed discussion see Ref. \cite{L-Z-77}). 
The results are equivalent to the ones presented above (see Table \ref{omn>2}).

The presence of a marginal operator both at Ising and at cubic fixed points
means that we should 
expect slowly decaying crossover effects that could 
produce systematic errors in Monte Carlo simulations and in the experimental 
investigations.

\subsection{Critical exponents}

The direct evaluation of the critical exponents for the two-dimensional $O(N)$
model leads to erroneous values because of the strong effect 
of nonanalytic terms.
In fact from the analysis of the four- and five-loop series of the 
$N=1$ model is found $\eta \sim 0.131$ \cite{os-00} and 
$\eta=0.13(7)$ \cite{L-Z-77} instead of the exactly known $\eta=1/4$.
The value of $\nu$ for the Ising universality class is quite 
good~($\nu\sim 0.96$ \cite{L-Z-77,os-00}). 
An our unpublished analysis of the $O(N)$ series~(already cited in 
Ref. \cite{cp-01}) shows that for higher values of $N$ the effect of 
nonanalyticities is very dangerous.
In fact, studying four-loop series,  we find $\eta=0.11(6)$ both for 
the $XY$ and Heisenberg model instead of $1/4$ and $0$. In the
same way for $\eta-\eta_t$ we find $1.18(5)$ and $1.36(4)$, for $N=2,\,3$,
instead of $2$. This latter systematic error brings to a finite value of $\nu$.
The above standard analysis applied to 
the $XY$ model with cubic anisotropy gives $\eta\sim 0.11$ along the
fixed-point line, and values ranging from $1.18$ to $0.97$ for $\eta-\eta_t$.

In order to reduce the effect of nonanalytic terms in the estimates of the 
critical exponents of the cubic model we adopt a new strategy.
We use a constrained analysis on the two variables series $\eta$ and 
$\eta-\eta_t$ fixing the values assumed at the fixed point on the axes
$\ub =0$ and $\vb =0$.
Explicitly, let us consider a generic $l$-loop series in $\ub$ and $\vb$
without constant term
\begin{equation}
R(\bar{u},\bar{v}) =\, \sum_{k=0}^l \sum_{h=0}^{l-k} R_{hk} \ub^h \vb^k=\, 
\sum_{k=1}^l R_{k}(z) \ub^k,
\end{equation}
of which we know the values $R(\ub^*,0)=a$ and $R(0,\vb^*)=b$.
We can rewrite the previous function in the form
\begin{equation}
R(\ub,\vb)=R(\ub,0)+R(0,\vb)+\ub \, \vb A_R(\ub,\vb)
\label{dec}
\end{equation}
where $\ub \, \vb A_R(\ub,\vb)$ is the difference between the original function
 $R(\ub,\vb)$ and
the value that it assumes on the axes. We resum the three functions
$R(\ub,0)\, ,\, R(0,\vb),\mbox{ and } A_R(\ub,\vb)$, appearing
in (\ref{dec}), in an independent way. In the case of $R(\ub,0)$ and $R(0,\vb)$
we adopt the standard method to resum one variable functions with some
constraints, as done for the $\epsilon$-expansion in Refs. 
\cite{L-Z-87,pv-98,cpv-99}. A full description of this method may be found 
in Ref. \cite{pv-98}.

\begin{figure}[t]
\centerline{\psfig{height=6truecm,width=8.6truecm,file=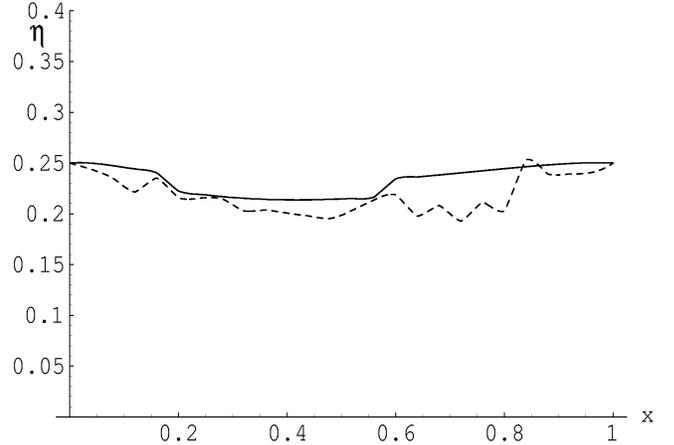}}
\caption{Values of $\eta$ with varying the parameter $x$ (\ref{zversusk}).
The straight line represents CM results and the dashed the PB ones.}
\label{etan2}
\end{figure}

\begin{figure}[b]
\centerline{\psfig{height=6truecm,width=8.6truecm,file=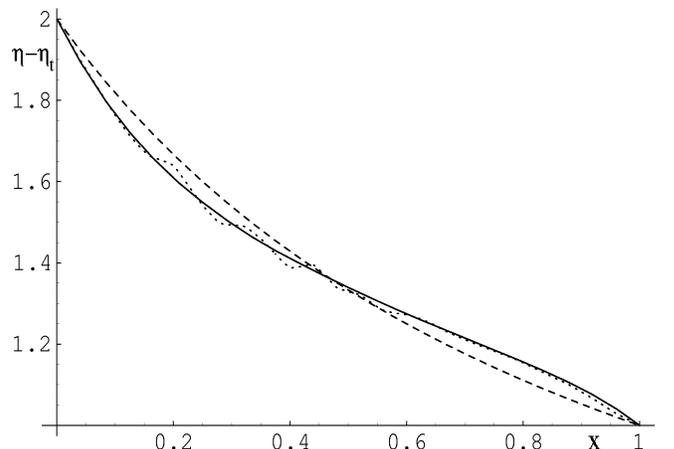}}
\caption{Values of $\eta-\eta_t$~(straight line CM and pointed line PB) 
with varying the parameter $x$ (\ref{zversusk}).
The dashed line is our conjecture (\ref{congettura2}).}
\label{ytn2}
\end{figure}

First of all we consider the peculiar case $N=2$. We evaluate the critical 
exponents on the straight line joining the Ising and the O(2) fixed points~(as 
in the case of the $\Omega$ matrix).
The results for the exponents $\eta$ and $\eta-\eta_t$ are presented in the 
Figs. \ref{etan2} and \ref{ytn2} respectively.
Within the precision of our calculation
we find an exponent $\eta$ assuming the constant value $1/4$ along 
the line as predicted in Ref. \cite{jkkn-77}~(the uncertainty 
is about $0.05$ for $x\sim 0.5$ and decreases near the borders). 
The exponent $\eta-\eta_t$ interpolates in a 
monotonic and continuous way from $1$ at
the Ising fixed point to $2$ at the $O(2)$ fixed point and so it is
compatible with the Kadanoff's conjecture identifying this line with
the dual line of the 8V model with the continuous varying exponent given
by (\ref{nuAT}). 

The parameter $y$ in equation (\ref{nuAT}) belongs to the range $[1,3/2]$
respectively from Ising to $O(2)$, instead  $x$ ranges between
$0$ and $1$ from $O(2)$ to Ising. So we try to identify
\begin{equation}
y=\frac{3-x}{2}.
\label{congettura}
\end{equation}
Using this assumption the exponent $\eta-\eta_t$ could be 
written in terms of $x$ as
\begin{equation}
\eta-\eta_t=\frac{2}{1+x}.
\label{congettura2}
\end{equation}
This curve is shown in Fig. \ref{ytn2} together with the numerical
values obtained from the resummation of the perturbative series: all
the curves are very close.
This result strongly
supports both  Kadanoff's conjecture about the nature of the line of
fixed points and the relation (\ref{congettura}) between the 
parameter $y$ of the 8V-model and $x$ of the $XY$ model with cubic anisotropy.

\begin{table}[b]
\caption{Critical exponents for $N \geq 3$ at the cubic fixed point.}
\label{expn>2}
\begin{tabular}{r|c|c}
N & $\eta$ [CM,PB] & $\eta-\eta_t$ [CM,PB]\\
\tableline \hline
8 & [0.26(2),0.25(2)]  & [1.20(15),1.15(15)]\\
4 & [0.23(4),0,20(6)]  & [1.22(15),1.18(20)]\\
3 & [0.17(8),0.15(9)]  & [1.39(20),1.23(20)]
\end{tabular}
\end{table}

The exponents at the cubic fixed point obtained using this constrained analysis 
for some $N\geq 3$ are reported in Table \ref{expn>2}. 
We find an $\eta$ exponent varying very slowly with $N$. 
Our estimates are also 
compatible with an $\eta$ constant and equal to $1/4$ for all $N$.
The estimates of $\eta-\eta_t$ are instead affected by a big error bar,
mainly due to the uncertainty of the cubic fixed points. We could
reduce the last error using the pseudo-$\epsilon$ expansion, but in this
way we will find again the systematic error of nonanalyticities that we cannot
control.

From Table \ref{expn>2} it is clear that when $N$ increases the exponents get
closer to the ones of the Ising model, that is the limit for 
$N \rightarrow \infty$, cf. (\ref{largen}). In this limit  the series 
$\eta_{\phi}$ reproduces order by order the one of the 
Ising model. For the difference $\eta-\eta_t$ the equality of the exponent 
is expected only at the fixed point. In fact it holds
\begin{equation}
\lim_{N\rightarrow \infty} \eta_t(\ub,\vb)=
-2 \ub + \eta_t^I (\vb)- \ub \, \vb^2 A(\vb)
\end{equation}
where $\eta_t^I (\vb)$ is the series of the Ising model,
and $A(\vb)$ a function that at four-loop reads $A(\vb)=0.1166-0.1157 \,\vb$.
At the cubic fixed point we obtain 
\begin{equation}
\eta_t-\eta_t^I =0.13(10)
\end{equation}
that is compatible with zero.

For $N=3$ the result for $\eta-\eta_t$ via scaling law leads to 
$\nu=1.3(3)$, so there are no fixed points~(stable or unstable) with the 
three-state Potts exponent $\nu=5/6$. 
In Sec. \ref{sec2c} we have stressed that experimental investigations 
\cite{pp-89} and Monte Carlo simulations \cite{cai-91,rj-93} in the 
adsorption of diatomic molecules on honeycomb lattice in the $p(2 \times 1)$ 
structure display the critical behavior of three-state Potts, although,
according to the standard classification \cite{ds-79}, this
system should be described by the $N=3$ cubic model.
It may be possible that the system undergoes a weak first-order phase 
transition in the face-cubic region with $v<0$, with effective exponents close
to the ones of the three-state Potts model which appears 
in the discrete face-centered model.
This point needs further studies for a full understanding of the phenomenon.

\section{Conclusions}

In the present paper we have studied the critical behavior of $N$-component
spin models with cubic anisotropy by applying the field-theoretic 
renormalization-group technique directly in two-dimensions.

We have firstly focused our attention to the $XY$ model~($N=2$) with
cubic anisotropy. We found that this model has a line of fixed points joining
the decoupled Ising and the $O(2)$. Along this line the critical exponent 
$\eta$
assumes the constant value $1/4$. The exponent $\nu$ runs from $1$, at Ising, 
to $\infty$, at $O(2)$, according to the standard expression of the 8V model
(\ref{nuAT}). This argument is based on a mapping of the $y$ parameter
of the 8V model onto the $x$ parameter of the Hamiltonian (\ref{Hphi4}):
\begin{equation}
x=3-2 y= \frac{2}{\pi} \arctan \frac{\vb^*}{\ub^*}.
\end{equation}
The predicted expression for $\nu$ (Eq.~(\ref{congettura2})) is in very good 
agreement with the resummation of perturbative series~(see Fig. \ref{ytn2}).
In this manner we relate the measured exponents to the strength of the 
anisotropy~(the parameter $x$) \cite{footnote2}. 
The $XY$ model with cubic anisotropy for $u<0$ could display the 
critical behavior of the Askin-Teller model with $y<1$, i.e., the region
on the dual line from the decoupled Ising fixed point~($y=1$) to the 
four-state Potts~($y=0$).
We tried to check if the line of fixed points continues for negative
value of $u$, but the results of the resummation for these values of the 
renormalized couplings are strongly oscillating. 

For all other values of $N$ we found one fixed
point in the region with $\ub,\vb >0$, that is the usual cubic fixed point, 
analytic continuation of one found in the $\epsilon$-expansion. 
The fixed point is marginally stable and unstable for perturbations with 
$v/u < v^*/u^*$ and $v/u > v^*/u^*$ respectively. The conjectured RG flow
diagram is sketched in Figure \ref{rgflow2}.
The estimates of this fixed point for several values 
of $N$ are reported in Table \ref{tfp} and the critical behavior is
characterized by the exponents of Table \ref{expn>2}.

In particular for the physical relevant case of $N=3$ at the cubic fixed point
we found 
\begin{equation}
\eta =0.17(8)\, , \qquad \nu =1.3(3).
\end{equation}
The value found for $\nu$ is  different from the one of the 
three-state Potts model.
This fact deserves further studies about the order-disorder transitions of some
adsorbed monolayers that should belong to the universality class
of the Heisenberg cubic model, but that experimentally display the exponents
of the three-state Potts model. One possible scenario is that the system 
undergoes a very weak first-order phase transition with 
effective exponents close to the ones of three-state Potts model.

\acknowledgements

We would like to thank Ettore Vicari and Andrea Pelissetto for a critical 
reading of this manuscript and many useful discussions. 
We also thank A.~I.~Sokolov and C.~Tanguy for useful correspondence.



\begin{references}
\bibitem{footnote0}
In this work we do not consider the quadratic term with cubic symmetry, 
that may exist for $N=2$, $\sum_{\mu =1}^2 (\partial_{\mu} \phi_{\mu})^2$ 
\cite{Aharony-73-3,Bruce-74,N-T-75,B-A-75,Aharony-76}. As we shall see, the 
quartic term in the Hamiltonian (\ref{Hphi4}) already introduces significant 
changes in the critical behavior.
\bibitem{footnote1}
It should be noted that in Refs. \cite{glb-90,bch-93} is found $\eta=0.38(12)$ 
and $\eta \sim 0.57$ respectively. 
These results are clearly different from the expected $\eta=1/4$.
\bibitem{footnote3}
In order to find more precise estimates of the cubic fixed point for $N\geq 3$ 
we also perform a constrained analysis of the function $\bar{\beta}_{\vb}$ 
fixing the zero for $\ub =0$ at $\vb =\bar{g}_I^*$. This analysis brings to
results substantially equivalent to the ones of Table \ref{tabexp}.
\bibitem{footnote2}
In order to relate the physical anisotropy to the experimentally measured 
exponents, it remains to understand
the relation between the ratio of the bare parameter $u_0$ and $v_0$~(that
are the physical couplings) and their fixed point values.
This could be obtained by a Monte Carlo simulation of the 
Hamiltonian (\ref{Hphi4}).
\end{references}
\end{document}